\documentclass[12pt]{article}

\usepackage{epsfig}
\usepackage{amssymb,amsmath}
\usepackage{slashed}

\newcommand{\bea}{\begin{eqnarray}}
\newcommand{\eea}{\end{eqnarray}}

 \makeatletter
\def\alt{\mathrel{\mathpalette\gl@align<}}
\def\agt{\mathrel{\mathpalette\gl@align>}}
\def\gl@align#1#2{\lower.6ex\vbox{\baselineskip\z@skip\lineskip\z@
\ialign{$\m@th#1\hfil##\hfil$\crcr#2\crcr\sim\crcr}}} \makeatother

\begin{document}
\begin{flushright}
 BA-07-36\\
 OSU-HEP-07-06
\end{flushright}
\vspace*{1.0cm}

\begin{center}
\baselineskip 20pt {\Large\bf Family Unification with SO(10)}
\vspace{1cm}

{\large K.S. Babu$^{a}$,  S.M. Barr$^{b}$ \vspace{.5cm} and  Ilia
Gogoladze$^{b,}$\footnote{ On  leave of absence from: Andronikashvili
Institute of Physics, GAS, 380077 Tbilisi, Georgia.
\\ \hspace*{0.5cm} }}

{\baselineskip 20pt \it $^a$Department of Physics, Oklahoma State
University, Stillwater,
OK 74078, USA \\
$^b$Bartol Research Institute, Department of Physics and Astronomy, \\
University of Delaware, Newark, DE 19716, USA \\
  } \vspace{.5cm}

\end{center}

\begin{abstract}
Unification based on the group $SO(10)^3 \times S_3$ is studied. Each family
has its own $SO(10)$ group, and the $S_3$ permutes the
three families and $SO(10)$ factors. This is the maximal local symmetry for
the known fermions. Family unification is achieved in the sense that all
known fermions are in a single irreducible multiplet of the symmetry.
The symmetry suppresses SUSY flavor changing effects by making all squarks and
sleptons degenerate in the symmetry limit. Doublet-triplet splitting can arise
simply, and non-trivial structure of the quark and lepton masses emerges from
the gauge symmetry, including the ``doubly lopsided" form.
\end{abstract}
\thispagestyle{empty}

\newpage

\addtocounter{page}{-1}

\baselineskip 18pt


In this paper we propose the idea of family unification based on the
group $SO(10) \times SO(10) \times SO(10) \times S_3$, where each
family of quarks and leptons transforms as a spinor under its own
$SO(10)$, and where the $S_3$  permutes the three families
and the three $SO(10)$ factors.

This structure has several interesting features. First, it may be
the only way to achieve family unification \cite{famunif}
satisfactorily in four space-time dimensions. (Attempts to unify
families in complex spinors of $SO(4n+2)$ groups have not resulted
in realistic models, since these spinors contain families and mirror
families when decomposed to the Standard Model symmetry
\cite{famunif}.  Family unification can occur in higher dimensions,
as in heterotic string theory \cite{string}. For family unification
in five, six and other dimensions, see \cite{famunifhighdim}.) Here
we define family unification to mean that all three families of
quarks and leptons including their right-handed neutrinos are
contained in a single irreducible representation of the unification
group, and that there is a single gauge coupling constant at high
scales. (This definition is broader than that often used, in which
the three families form an irreducible representation of a {\it
simple} group.) It should be noted that in the scheme we propose
here the families are in a reducible representation of $SO(10)^3$,
namely $\{(16, 1, 1) + (1,16,1) + (1, 1, 16)\}$, but these form an
{\it irreducible} multiplet of the full unification group that
includes the $S_3$ factor.

Second, the group $SO(10) \times SO(10) \times SO(10) \times S_3$ is
the largest that can be gauged with 48 fermions forming a complex
but anomaly-free set of representations. In that sense, it is the
``maximal local symmetry" of the known quarks and leptons, including
the right-handed neutrinos. (In \cite{maxlocalsym}, the definition
of maximal local symmetry also included the condition that the group
be simple. By that more restrictive definition, the maximal local
symmetry of 48 fermions would be just $SO(10)$.)  It is easy to see
that the $S_3$ factor is anomaly free.  $S_3$ makes the gauge
couplings of all $SO(10)$ groups equal.  As a result, the instanton
effects of the  three $SO(10)$ groups will also be $S_3$--invariant,
proving that it is anomaly free.  (The cyclic permutation group
$Z_3$, which is often considered in such contexts is a subgroup of
this $S_3$. Incidently, in the model we present it might appear that
it is possible to gauge additional $U(1)$ factors, where under these
$U(1)$'s the fermions are rotated into themselves by a phase factor
and the gauge bosons are invariant.  However, the only anomaly free
part of these $U(1)$'s  are the $Z_4$ centers of the $SO(10)$
groups.  It is interesting that the $S_3$ does not commute with
these $Z_4$, and other $SO(10)$, transformations.)

Third, the unification of all the known quarks and leptons in a
single irreducible multiplet of a local group suppresses SUSY flavor
changing processes by making all squark and slepton masses exactly
degenerate at the unification scale.

Fourth, the problem of ``doublet-triplet splitting" can be solved
more easily with this group than in ordinary $SO(10)$ unification by
means of the Dimopoulos-Wilczek mechanism \cite{dw} also known as
the ``missing VEV mechanism". That the Dimopoulos-Wiczek mechanism
is straightforward to implement in product groups like $SU(5) \times
SU(5)$ and $SO(10) \times SO(10)$ was pointed out in
\cite{wali,dwproduct,witten}.  The stability of the VEV structure
however is nontrivial to achieve in $SO(10) \times SO(10)$ models,
as that requires rank reduction of the diagonal $SO(10)$
\cite{so10}.  As will be shown, the present framework resolves this
issue very neatly.

And, finally, the ``vertical" group $SO(10)^3$ is also in a sense a ``family
group", since the three families transform differently under any one of
the $SO(10)$ groups. As will be seen later, highly non-trivial
patterns emerge in the mass matrices of the quarks and leptons due primarily
to the symmetry $SO(10)^3$, though $S_3$ also plays a role.
Some of the patterns that emerge almost automatically in this framework
have already been proposed in the literature on
purely phenomenological grounds, such as the ``doubly lopsided"
structure \cite{doublylopsided}.

We will describe a supersymmetric $SO(10)^3 \times S_3$ model that
illustrates some of the possibilities of the idea. The quarks and
leptons are in the representation $\{(16, 1, 1) + (1, 16, 1) + (1,
1, 16)\}$, which we shall denote $(16, 1, 1) + cyclic$ for short.
The Higgs doublets of the Standard Model are contained in the
``fundamental" Higgs multiplet $(10, 1, 1) + cyclic$. Two kinds of
Higgs multiplets are needed to do breaking of $SU(3)^3 \times S_3$
to the Standard Model and give superheavy mass to the right-handed
neutrinos. We take these to be the ``bifundamental" Higgs multiplet
$(10, 10, 1) + cyclic$ and the ``bispinor" Higgs multiplet $(16, 16,
1) + cyclic$ (plus the conjugate bispinor multiplet $(\overline{16},
\overline{16}, 1)$).

The Standard Model group is contained within the ``diagonal
$SO(10)$" of the three factor $SO(10)$ groups. Under this diagonal
$SO(10)$ subgroup, the bifundamentals contain the representations
${\bf 1} + {\bf 45} + {\bf 54}$, while the bispinors contain ${\bf
10} + {\bf 126} + {\bf 120}$. It is well-known that a Higgs field in
the bifundamental representation of a group $G \times G$ can break
it to the diagonal subgroup $G$. Similarly, a set of bifundamentals
can break $G \times G \times G$ to the diagonal subgroup of the
three factor groups. In our  model, there is a minimum of the
scalar potential where the vacuum expectation values (VEVs) of the
bifundamentals break $SO(10)^3$ all the way down to a diagonal
$SU(3)_c \times SU(2)_L \times U(1) \times U(1)$, as will be seen.
The bispinors, whose VEVs give mass to the right-handed neutrinos,
break the extra $U(1)$ to give the Standard Model group.

The quark and lepton multiplets will be denoted $\psi_a$, $a =
1,2,3$, where $\psi_1 \equiv (16, 1, 1)$, $\psi_2 \equiv (1,16,1)$,
and $\psi_3 \equiv (1,1,16)$. The fundamental Higgs fields will be
denoted $H_a$, where $H_1 \equiv (10, 1, 1)$, etc. The bifundamental
Higgs fields will be denoted $\Omega_{ab}$, where $\Omega_{12}
\equiv (10,10,1)$, etc. A second set of bifundamentals will also be
needed, and will be denoted $\Omega'_{ab}$. And the bispinors will
be denoted $\Delta_{ab}$ and $\overline{\Delta}_{ab}$, where
$\Delta_{12} \equiv (16, 16, 1)$ and $\overline{\Delta}_{12} \equiv
(\overline{16}, \overline{16}, 1)$, etc. The indices $a$ and $b$ are
not $SO(10)$ indices (which we suppress) but merely labels that
indicate which $SO(10)$ groups the multiplets transform
non-trivially under. These labels are permuted under the
$S_3$ group.

Only two renormalizable terms are allowed by the gauge symmetry in
the Yukawa superpotential of the quarks and leptons, namely
\begin{equation}
W_{Yuk} = Y (\psi_1 \psi_1 H_1 + \psi_2 \psi_2 H_2 + \psi_3 \psi_3
H_3) + Y' (\psi_1 \psi_2 \overline{\Delta}_{12} + \psi_2 \psi_3
\overline{\Delta}_{23} + \psi_3 \psi_1 \overline{\Delta}_{31}).
\end{equation}

The Higgs superpotential can be written in an obvious notation as
$W_{Higgs} = W_H + W_{\Omega} + W_{\Delta} + W_{H \Omega} +
W_{\Omega \Delta} + W_{H \Delta}$. Let us focus first on
$W_{\Omega}$. If there is only one set of bifundamentals
$\Omega_{ab}$, then the most general renormalizable form of
$W_{\Omega}$ consistent with symmetry is
\begin{equation}
W_{\Omega} = \frac{1}{2} M (tr \Omega_{12} \Omega_{21} + tr
\Omega_{23} \Omega_{32} + tr \Omega_{31} \Omega_{13}) + \lambda (tr
\Omega_{12} \Omega_{23} \Omega_{31}).
\end{equation}

\noindent The order of labels $ab$ on $\Omega_{ab}$ is significant.
If $(\Omega_{ab})^{ij}$ is the $ij$ element of the $10 \times 10$
matrix $\Omega_{ab}$, then the row index $i$ belongs to the group
$SO(10)_a$ and the column index $j$ belongs to the group $SO(10)_b$.
This superpotential gives rise to the equations of motion
$\Omega_{21} = (\lambda/M) \Omega_{23} \Omega_{31}$, $\Omega_{32} =
(\lambda/M) \Omega_{31} \Omega_{12}$, and $\Omega_{13} = (\lambda/M)
\Omega_{12} \Omega_{23}$, which, of course, are $S_3$ permuted
versions of each other. Doublet-triplet splitting by means of the
Dimopoulos-Wilczek mechanism would require that the VEV of at least
one of these bifundamentals had the form $\langle \Omega \rangle =$
diag $(a~a~a~0~0) \otimes i \tau_2$ (corresponding to the generator
$B-L$ of the diagonal $SO(10)$ and the generator diag $(a~a~a~0~0)$ of
the diagonal $U(5)$). However, it is easily seen from the three
equations of motion, that if any one of the $\Omega_{ab}$ has a VEV
of this form, all three of them would have a similar form (in some
basis), i.e. would vanish in the lower $4 \times 4$ block and therefore a
subgroup $SO(4)^3 = (SU(2)_L \times SU(2)_R)^3$ would remain
unbroken by the bifundamentals. And even after breaking by the
bispinor Higgs VEVs there would still remain an unbroken
$(SU(2)_L)^3$. Thus the form in Eq. (2) is too simple to give the
bifundamental VEVs the Dimopoulos-Wilczek form and also break the
symmetry down to the Standard Model group at low energies.

Satisfactory breaking can happen if there are two sets of
bifundamentals, $\Omega_{ab}$ and $\Omega'_{ab}$, where
$\Omega'_{ab}$ is odd under a $Z_2$ parity and $\Omega_{ab}$ is
even. Then the most general renormalizable form of $W_{\Omega}$ is
\begin{equation}
\begin{array}{ccl}
W_{\Omega} & = & \frac{1}{2} M (tr \Omega_{12} \Omega_{21} + tr
\Omega_{23} \Omega_{32}
+ tr \Omega_{31} \Omega_{13})
+  \frac{1}{2} M' (tr \Omega'_{12} \Omega'_{21} + tr \Omega'_{23}
\Omega'_{32}
+ tr \Omega'_{31} \Omega'_{13}) \\
& & \\
& + & \lambda (tr \Omega_{12} \Omega_{23} \Omega_{31})
+ \lambda' (tr \Omega_{12} \Omega'_{23} \Omega'_{31} + tr
\Omega'_{12} \Omega_{23} \Omega'_{31} + tr \Omega'_{12} \Omega'_{23}
\Omega_{31}) \end{array}
\end{equation}

\noindent The equations of motion then become: $\Omega_{ba} =
(\lambda/M) \Omega_{bc} \Omega_{ca} + (\lambda'/M) \Omega'_{bc}
\Omega'_{ca}$ and $\Omega'_{ba} = (\lambda'/M') \times$ \newline  $(\Omega'_{bc}
\Omega_{ca} + \Omega_{bc} \Omega'_{ca})$. These equations have many
interesting solutions. The one that seems phenomenologically most
interesting has VEVs of the following form:

\begin{equation}
\begin{array}{ll}
\Omega_{12} = \left( \begin{array}{cc} A & 0 \\ 0 & B \end{array}
\right), & \Omega_{23} = \Omega_{31} = \left( \begin{array}{cc} A &
0 \\ 0 & 0 \end{array} \right),
\\ & \\
\Omega'_{12} = \left( \begin{array}{cc} A' & 0 \\ 0 & 0 \end{array}
\right), & \Omega'_{23} = \Omega'_{31} = \left( \begin{array}{cc} A'
& 0 \\ 0 & B' \end{array} \right),
\end{array}
\end{equation}

\noindent where $A$ and $A'$ are $6 \times 6$ matrices and $B$
and $B'$ are $4 \times 4$  matrices, given by
\begin{equation}
A = a I_6, \;\;\; A' = a' I_3 \otimes i \tau_2, \;\;\; B = b I_4,
\;\;\; B' = b' I_2 \otimes i \tau_2,
\end{equation}

\noindent and where $a = \frac{M'}{2\lambda'}$, $a' = \sqrt{\frac{M
M'}{2 \lambda^{\prime 2}} + \frac{\lambda M^{\prime 2}}{4
\lambda^{\prime 3}}}$, $b= - \frac{M'}{\lambda'}$, and $b' =
\frac{\sqrt{-M M'}}{\lambda'}$. At this minimum, the bifundamentals
break $SO(10)^3$ down to a diagonal $SU(3) \times SU(2) \times U(1)
\times U(1)$.

To illustrate some of the possibilities, we mention a few other
solutions of the many that exist: (i) One can have the same form as in Eqs.
(4) and (5) but with $A' = a' I_6$, $a' = \sqrt{\frac{M M'}{2
\lambda^{\prime 2}} - \frac{\lambda M^{\prime 2}}{4 \lambda^{\prime
3}}}$. This breaks $SO(10)^3$ down to $SU(4) \times SU(2) \times
U(1) \times U(1)$. (ii) One can have the same form as in Eqs. (4)
and (5) but with $B' = b' I_4$, $b' = \frac{\sqrt{+M
M'}}{\lambda'}$. This breaks down to $SU(3) \times SU(2) \times
SU(2) \times U(1)$. (iii) One can have the same form as in Eqs. (4)
and (5) but with {\it both} the substitutions of cases (i) and (ii).
This would break the group only down to the Pati-Salam group $SU(4)
\times SU(2) \times SU(2)$. (iv) There are solutions where all three
of the $\Omega_{ab}$ have the form that $\Omega_{12}$ has in Eq. (4)
and all three of the $\Omega'_{ab}$ have the form that
$\Omega'_{12}$ has in Eq. (4). (v) There are solutions where the
matrices have different forms than shown in Eq. (4), for example
having different rank than 4, 6, or 10. Some of the unbroken groups
that can result are $SU(5) \times U(1)$, $SU(4) \times U(1) \times
U(1)$, $SO(8) \times SO(2)$. A complete analysis of all the minima
would be rather lengthy.

The form in Eq. (4) is a useful one for the purposes of
doublet-triplet splitting, as it can lead to a single pair of Higgs
doublets being light. To see this, consider $W_{H \Omega}$, whose
most general renormalizable form consistent with symmetry is
\begin{equation}
W_{H \Omega} = \lambda_{H \Omega} ( H_1 \Omega_{12} H_2 + H_2
\Omega_{23} H_3 + H_3 \Omega_{31} H_1).
\end{equation}

\noindent Similar terms with $\Omega'_{ab}$ are ruled out by the
$Z_2$ parity. If the explicit mass term $(H_1 H_1 + H_2 H_2 + H_3
H_3)$ is forbidden (or suppressed to be of order
the weak scale) by symmetry (for example a softly broken discrete
symmetry or an $R$ symmetry) then the mass matrix
of the color-triplets and weak-doublets in $H_a$ have the form

\begin{equation}
M_3 = \left( \begin{array}{ccc} 0 & a & a \\ a & 0 & a \\ a & a & 0
\end{array} \right), \;\;\; M_2 = \left( \begin{array}{ccc}
0 & b & 0 \\ b & 0 & 0 \\ 0 & 0 & 0 \end{array} \right).
\end{equation}

\noindent It is apparent that only a single pair of doublet Higgs
fields (namely those in $H_3$) remain light, as needed for
gauge-coupling unification. Therefore, only $H_3$ will get a
weak-interaction-breaking VEV, and because of the form of the Yukawa
superpotential given in Eq. (1) only the third family of quarks and
leptons will get mass. Below it will be shown that higher-dimension
operators can generate other entries in the quark and lepton mass
matrices, allowing non-zero (but presumably smaller) masses for the
other families and CKM mixing. It is interesting that the $SO(10)^3
\times S_3$ symmetry and a choice of minimum consistent with a
single pair of light Higgs doublets leads to a natural hierarchy
wherein one family is heavier than the others.

In order to give mass to the right-handed neutrinos the bispinors
$\overline{\Delta}_{ab}$ must receive non-vanishing VEVs such that
both spinors of the bispinor point in the Standard-Model-singlet
direction.  One possible superpotential which achieves this is given below.
\begin{eqnarray}
W_\Delta = \lambda_\Delta\{(\Delta_{12} \overline{\Delta}_{12}-M^2) S_{12} +
(\Delta_{23} \overline{\Delta}_{23} - M^2) S_{23}+
(\Delta_{31} \overline{\Delta}_{31}-M^2) S_{31}\}~,
\end{eqnarray}
where $S_{ab}$ are $SO(10)$ singlets.  This superpotential admits
a solution where all three $\Delta_{ab}$'s and $\overline{\Delta}_{ab}$'s
have equal VEVs along
their respective Standard Model singlet directions.  Higher dimensional
operators of the form $(\Delta \overline{\Delta})^2/M_*$ will have to
be introduced to give masses to all pseudo-Goldston bosons from these
fields.

An interesting feature of this VEV structure is that from Eq. (1),
it will generate a Majorana right--handed neutrino mass matrix
which has equal entries in the off-diagonals, and zero entries along
the diagonals (as in $M_3$ of Eq. (7)).  That will result in two
degnerate $\nu^c$ fields, which may be relevant for resonant leptogenesis.

There are only two renormalizable terms allowed in $W_{\Omega
\Delta}$ by symmetry, namely $(\Delta_{ab} \Delta_{ab} \Omega_{ab} +
cyclic)$ and $(\overline{\Delta}_{ab} \overline{\Delta}_{ab}
\Omega_{ab} + cyclic)$. These terms give sufficient coupling between
the bifundamental Higgs sector and the bispinor Higgs sector to
prevent any uneaten goldstone bosons. (With insufficient coupling
between two kinds of Higgs, goldstone bosons can arise that
correspond to relative rotations of their VEVs.)  It is in this
regard that the present model fares better than the usual $SO(10)$
models, where new mechanisms should come in to stabilize the VEV
structure \cite{br}.

Turning to higher-dimension operators, one finds that there are only
a few quartic operators allowed by the $SO(10)^3 \times S_3$
symmetry in the Higgs superpotential. Some of these (such as $(H_a^2
\Omega_{bc}^2 + cyclic))$ can be constructed by multiplying pairs of
the invariant quadratic operators $H_a^2$, $\Omega_{ab}^2$, and
$\Delta_{ab} \overline{\Delta}_{ab}$ (or by taking such products and
contracting the gauge indices differently). In addition, there are
the following five types of invariant quartic operators: $O_1 \equiv
(H_a H_b \Omega_{ca} \Omega_{cb} + cyclic)$; $O_2 \equiv (H_a H_b
\Delta_{ab}^2 + cyclic)$ and $\overline{O}_2 \equiv (H_a H_b
\overline{\Delta}_{ab}^2 + cyclic)$; $O_3 \equiv (\Omega_{ab}
\Omega_{ac} \Delta_{bc}^2 + cyclic)$ and $\overline{O}_3 \equiv
(\Omega_{ab} \Omega_{ac} \overline{\Delta}_{bc}^2 + cyclic)$; $O_4
\equiv (\Delta_{ab}^4 + cyclic)$ and $\overline{O}_4 \equiv
(\overline{\Delta}_{ab}^4 + cyclic)$; and $O_5 \equiv (\Delta_{ab}
\Delta_{bc} \overline{\Delta}_{ca} H_b + cyclic)$ and
$\overline{O}_5 \equiv (\overline{\Delta}_{ab}
\overline{\Delta}_{bc} \Delta_{ca} H_b + cyclic)$. The operators of
type $O_1$, $O_3$, and $\overline{O}_3$ can exist with the product
of bifundamentals being $\Omega \Omega$, $\Omega \Omega'$, or
$\Omega' \Omega'$, as far as the symmetry $SO(10)^3 \times S_3$ is
concerned; which of these is actually allowed in the superpotential
depends on the $Z_2$ parity assignments of the fields.

There are only two types of quartic operators allowed by $SO(10)^3
\times S_3$ in the Yukawa superpotential, namely $O_{Y1} \equiv
(\psi_a \psi_a H_b \Omega_{ab} + cyclic)$, $O_{Y2} \equiv (\psi_a
\psi_b \Omega_{ab} \Delta_{ab} + cyclic)$. Again, in both cases
$SO(10)^3 \times S_3$ permits such operators with either $\Omega$ or
$\Omega'$, whereas some will not be allowed by $Z_2$ parity.

These quartic operators, which may be induced either by Planck-scale
physics or by integrating out fields at the unification scale, have
interesting consequences.  Consider first the operator $O_5$, which
written out is $\Delta_{12} \Delta_{23} \overline{\Delta}_{31} H_2 +
\Delta_{23} \Delta_{31} \overline{\Delta}_{12} H_3 + \Delta_{31}
\Delta_{12} \overline{\Delta}_{23} H_1$. The second term, which
involves $H_3$, is interesting because it induces weak-scale
$SU(2)_L$-breaking VEVs in both $\Delta_{23}$ and $\Delta_{31}$.

This happens as follows. If we write out this term in $SO(10)^3$
notation, it has the form $\Delta_{23} \Delta_{31}
\overline{\Delta}_{12} H_3 = (1, 16, 16) (16, 1, 16) (\overline{16},
\overline{16}, 1) (1, 1, 10)$. Using $[SU(5) \times U(1)]^3$
notation \cite{slansky}, the first factor ($\Delta_{23}$) has a
superlarge VEV in the $(1^0, 1^5, 1^5)$ direction, the second factor
($\Delta_{31}$) has a superlarge VEV in the $(1^5, 1^0, 1^5)$
direction, the third factor ($\overline{\Delta}_{12}$) has a
superlarge VEV in the $(1^{-5}, 1^{-5}, 1^0)$ direction, and the
last factor ($H_3$) has a weak-scale VEV in the $(1^0, 1^0, 5^{-2})$
direction. Therefore, there is effectively a linear term for the
$(1^0, 1^5, \overline{5}^{-3})$ component of $\Delta_{23}$ that
arises from this product: $(1^0, 1^5, \overline{5}^{-3}) \cdot
\langle (1^5, 1^0, 1^5) \rangle \langle (1^{-5}, 1^{-5}, 1^0)
\rangle \langle (1^0, 1^0, 5^{-2}) \rangle$. It is easy to see that
this will induce a weak-scale VEV in this component. So we may write
$\langle \Delta_{23}(1^0, 1^5, \overline{5}^{-3}) \rangle \sim M_W$.
Similarly there is a linear term for the $(1^5, 1^0,
\overline{5}^{-3})$ component of the $\Delta_{31}$ coming from the
product $\langle (1^0, 1^5, 1^5) \rangle \cdot (1^5, 1^0,
\overline{5}^{-3}) \cdot \langle (1^{-5}, 1^{-5}, 1^0) \rangle
\langle (1^0, 1^0, 5^{-2}) \rangle$. So we may write $\langle
\Delta_{31}(1^5, 1^0, \overline{5}^{-3}) \rangle \sim M_W$.

These weak-scale VEVs in $\Delta_{23}$ and $\Delta_{31}$ are
interesting, in turn, because they can contribute to quark and
lepton masses if the operator $O_{Y2}$ is present (in either its
$\Omega$ or its $\Omega'$ form). If one examines the term $\psi_2
\psi_3 \Omega'_{23} \Delta_{23} = (1, 16, 1) (1, 1, 16) (1, 10, 10)
(1, 16, 16)$, ones sees that it contains $(1^0, \overline{5}^{-3},
1^0)$ $(1^0, 1^0, 10^1) \langle (1^0, 5^{-2}, \overline{5}^2)
\rangle \langle (1^0, 1^5, \overline{5}^{-3}) \rangle$. In $SU(5)$
language, this is a contribution to a term of the form $\overline{5}_2
10_3 \langle \overline{5}_H \rangle$, i.e. the $10$ of the
third family times the $\overline{5}$ of the second family.
Therefore this operator gives a 23 element of the charged-lepton mass matrix
$M_L$ and a 32 element of the down-quark mass
matrix $M_D$.  It does not contribute to any other components of these
matrices, and it does not contribute to the up-quark mass matrix.
This is exactly the kind of entry that is needed in the so-called
``lopsided" mass matrix models \cite{lopsided}. If such entries are
comparable to
the 33 elements of $M_D$ and $M_L$, then they explain the fact that
the 2-3 mixing angle is large for the left-handed leptons (i.e.
$U_{\mu 3} \sim 1$) but small for the left-handed quarks (i.e.
$V_{cb} \ll 1$). If the relevant quartic terms arise from
integrating out fields with mass of order $M_{GUT}$ rather than
$M_{P \ell}$, there is no reason that these lopsided mass matrix
elements necessarily have to be smaller than the 33 elements, even
though the latter arise from cubic terms. (Even with quartic terms
generating the 23 and 13 entries, they may be comparable to the 33 entry if $\tan\beta$ is small.)
It should be noted that in
most published lopsided models the operator that gives the lopsided
entries $(M_D)_{32}$ and $(M_L)_{23}$ is such that these entries are
equal in magnitude. It is important that they be at least
approximately equal to reproduce the well-known prediction that at
the unification scale $m_b = m_{\tau}$. Here, the
operator $\psi_2 \psi_3
\Omega'_{23} \Delta_{23}$ does not make these entries equal but
gives them a ratio $(M_D)_{32}/(M_L)_{23} = a'/b'$, whose value
depends on the parameters in the Higgs superpotential. (See Eq.
(5).) (If, instead of $\Omega'$ in this operator there were $\Omega$,
then the contribution to $M_L$ would vanish.)

In the same way, it is easy to see that the weak-scale VEV of
$\Delta_{31}$ can generate contributions to the 31 element of $M_D$
and the 13 element of $M_L$. If these too are comparable to the
magnitudes of the 33 elements, then a so-called ``doubly lopsided"
model results \cite{doublylopsided}. As has been explained in the
literature, such models
can account for the so-called ``bi-large" pattern of neutrino mixing
in a very simple way, and have other attractive features as well.

One  might expect the transposes of these lopsided mass matrix
elements also to be induced by these quartic terms (e.g. the 23
element of $M_D$ in addition to the 32 element, etc.). However, they
are not. Nor are any off-diagonal elements of the up quark-mass
matrices, induced until one takes into account terms higher-order
than quartic.  This may well be related to the stronger mass
hierarchy observed among the up-type quarks.  Indeed, in lopsided
models, it is precisely the absence of large lopsided terms in $M_U$
that is responsible for this. It is noteworthy that in the lopsided
models published in the literature the placement of the lopsided
entries (for example that they appear in the 32 elements but not the
12 elements, say) is to some extent contrived with an eye to
reproducing the observed pattern of masses and mixings. Here, it is
largely dictated by the $SO(10)^3 \times Z_3$ symmetry of the theory
(and by the requirement that only one pair of Higgs doublets
remains light).

The operators $O_{Y1}$ can also play an important role. The term
$\psi_1 \psi_1 H_3 \Omega'_{13}$ and the term $\psi_2 \psi_2 H_3
\Omega'_{23}$ give 11 and 22 elements respectively to all the quark
and lepton mass matrices. Note
that the two terms $\psi_1 \psi_1 H_3 \Omega'_{13}$ and $\psi_2
\psi_2 H_3 \Omega'_{23}$ are related to each other by $S_3$. However,
if $S_3$ is broken spontaneously, either completely or down to a $Z_3$
subgroup, the equality of the 11 and 22 elements need not hold.

The following shows at what level various elements in the quark and
lepton mass matrices arise. A ``3" means that such an element can
arise even if only terms cubic and lower exist in $W$; a ``4" means
that such an element can arise only if quartic (or higher) terms are
present, and a ``5" means
that such an element can arise only if quintic (or higher) terms are
present.

\begin{equation}
M_U = \left( \begin{array}{ccc} 4 & 5 & 5 \\ 5 & 4 & 5 \\ 5 & 5 & 3
\end{array} \right), \;\; M_D = \left( \begin{array}{ccc} 4 & 5 & 5
\\ 5 & 4 & 5 \\ 4 & 4 & 3 \end{array} \right), \;\; M_L = \left(
\begin{array}{ccc} 4 & 5 & 4 \\ 5 & 4 & 4 \\ 5 & 5 & 3 \end{array}
\right).
\end{equation}

The entries that are labelled ``5" arise in a somewhat non-trivial way.
Consider, for instance, the 12 and 21 elements of the mass matrices.
The quintic term $\Delta_{12} \Delta_{13} \overline{\Delta}_{23}
\Omega_{13} H_3$ induces a weak-scale VEV in the component
$\Delta_{12} (\overline{5}^{-3}, 1^5, 1^0)$ through the product
$(\overline{5}^{-3}, 1^5, 1^0) \cdot \langle ( 1^5, 1^0, 1^5)\rangle$
$\langle (1^0, 1^{-5} 1^{-5}) \rangle \langle (5^{-2}, 1^0,
\overline{5}^2) \rangle \langle (1^0, 1^0, 5^{-2}) \rangle$.
This VEV then gives $(M_D)_{12}$ and
$(M_L)_{21}$ via the quartic Yukawa operator
$\psi_1 \psi_2 \Omega_{12} \Delta_{12}$ as follows:
$(10^1, 1^0, 1^0) (1^0, \overline{5}^{-3}, 1^0)
 \langle (\overline{5}^2, 5^{-2}, 1^0) \rangle
\langle (\overline{5}^{-3}, 1^5, 1^0) \rangle$. The transposed
elements $(M_D)_{21}$ and $(M_L)_{12}$
arise in a completely analogous way. (Just interchange the labels
1 and 2 everywhere in the preceding discussion.)
The 12 and 21 elements of the
up quark mass matrix $M_U$ can come from a quintic Yukawa operator
$\psi_1 \psi_2  \Omega_{12} \Omega_{12} \overline{\Delta}_{12}$
from the term $(10^1, 1^0, 1^0) (1^0, 10^1, 1^0)
 \langle (5^{-2}, \overline{5}^2,
1^0) \rangle \langle (5^{-2}, \overline{5}^2,
1^0) \rangle \langle (5^3, 1^{-5}, 1^0) \rangle$.

The same quintic term $\Delta_{12} \Delta_{13}
\overline{\Delta}_{23} \Omega_{13} H_3$, induces a weak-scale VEV
for $\Delta_{13} (\overline{5}^{-3}, 1^0, 1^5)$ through the product
$\langle (1^5, 1^5, 1^0) \rangle \cdot (\overline{5}^{-3}, 1^0, 1^5)
\cdot \langle (1^0, 1^{-5}, 1^{-5}) \rangle \langle (5^{-2}, 1^0,
\overline{5}^2) \rangle \langle (1^0, 1^0, 5^{-2}) \rangle$.  This
VEV then induces $(M_D)_{13}$ and $(M_L)_{31}$ through the quartic
term that was mentioned before as giving $(M_D)_{31}$ and
$(M_L)_{13}$, namely $\psi_1 \psi_3 \Omega'_{13} \Delta_{13}$. In
particular, this contains the product $(10^1, 1^0, 1^0) (1^0, 1^0,
\overline{5}^{-3}) \langle (\overline{5}^2, 1^0, 5^{-2}) \rangle
\langle (\overline{5}^{-3}, 1^0, 1^5) \rangle$.  The entries
$(M_D)_{23}$ and $(M_L)_{32}$ arise in a similar way. (Just
interchange the indices 1 and 2 in the foregoing discussion.)

What remains is to show that the 13, 31, 23, and 32 elements of
$M_U$ can arise from quintic terms. The elements $(M_U)_{13}$
and $(M_U)_{31}$ arise from the quintic term
$\psi_1 \psi_3 \Omega_{13} \Omega_{13} \overline{\Delta}_{13}$,
which contains the product
$(10^1, 1^0, 1^0) (1^0, 1^0, 10^1)
 \langle (\overline{5}^2, 1^0,
5^{-2}) \rangle  \langle (\overline{5}^2, 1^0,
5^{-2}) \rangle \langle (1^{-5}, 1^0, 5^3) \rangle$.
The elements $(M_U)_{12}$ and $(M_U)_{21}$
arise in a similar way.

One sees, then, that the requirements of $SO(10)^3 \times S_3$
symmetry imply that the quark and lepton mass matrices have
a non-trivial structure that contains several
promising features: (i) a hierarchy among the mass matrix elements,
(ii) only one family obtaining mass at lowest order, (iii) a
qualitative difference between the up quark mass matrix and the
other mass matrices (in particular some of the elements of $M_U$
arise at higher order than the corresponding elements
of $M_D$ and $M_L$, which is perhaps related to the stronger
hierarchy observed among the up-type quarks); (iv) relatively
large off-diagonal elements in the third row of $M_D$ and
third column of $M_L$, i.e. the ``doubly lopsided" pattern that is
known to explain in a simple way the bilarge pattern of
neutrino mixing; (v) ``Clebsches" in certain elements of
$M_D$ and $M_L$ that may allow an explanation of the
well-known Georgi-Jarlskog relations. Still,
the construction of a complete model with fully realistic
quark and lepton mass matrices has not been attempted here.

There are several issues that would have to be faced in constructing
a fully realistic model based on $SO(10)^3 \times S_3$. The most
difficult would be proton decay via the $d=5$ operators that arise
from the exchange of colored Higgsinos. The simple structure in Eq.
(7) leads to no suppression of such decay amplitudes. It seems
likely, however, that with more than two types of bifundamental
Higgs fields adequate suppression may be achieved.  Another issue is
the existence of Landau poles above the unification scale (i.e. the
$SO(10)^3 \times S_3$ scale) due to the large number of fields in
the bispinor and bifundamental Higgs mutiplets.  These issues
require further study.

\section*{Acknowledgments}

This work is supported in part by the DOE Grant \# DE-FG03-98ER-41076 (KSB)
and by DE-FG02-91ER40626 (SMB and IG).



\begin{thebibliography}{99}



\bibitem{famunif}

M. Gell-Mann, P. Ramond and R. Slansky, in {\it Supergravity},
edited by P. Van Nieuwenhuizen and D.Z. Freedman (North Holland,
Amsterdam, 1979); \newline
F.~Wilczek and A.~Zee,  Phys.\ Rev.\  D {\bf 25} (1982)  553.

\bibitem{string}
D.~J.~Gross, J.~A.~Harvey, E.~J.~Martinec and R.~Rohm,
Phys.\ Rev.\ Lett.\  {\bf 54} (1985) 502.


\bibitem{famunifhighdim}
  K.~S.~Babu, S.~M.~Barr and Bum-seok Kyae,
  Phys.\ Rev.\ D {\bf 65} (2002) 115008,
  [arXiv:hep-ph/0202178];
I.~Gogoladze, C.~A.~Lee, Y.~Mimura and Q.~Shafi,
  Phys.\ Lett.\  B {\bf 649}, 212 (2007),
  [arXiv:hep-ph/0703107];
   J.~E.~Kim,
  arXiv:0707.3292 [hep-ph].

\bibitem{maxlocalsym}
  A.~Zee, Phys.\ Lett. \ B {\bf 99} (1981) 110.

\bibitem{dw}
  S.~Dimopoulos and F.~Wilczek,
Print-81-0600 (SANTA BARBARA);\\
K.~S.~Babu and S.~M.~Barr,
  Phys.\ Rev.\ D {\bf 48} (1993) 5354,
  [arXiv:hep-ph/9306242].


\bibitem{wali}
 A.~Davidson and K.~C.~Wali,
  Phys.\ Rev.\ Lett.\  {\bf 59}, 393 (1987);
P.~L.~Cho,  Phys.\ Rev.\  D {\bf 48}, 5331 (1993),
  [arXiv:hep-ph/9304223];
 R.~N.~Mohapatra,
  Phys.\ Rev.\  D {\bf 54} (1996) 5728.

\bibitem{dwproduct}
  R.~Barbieri, G.~Dvali and A.~Strumia, Phys.\ Lett.\ B
  {\bf 333} (1994) 79, [arXiv:hep-ph/9404278];
  S.~M.~Barr, Phys.\ Rev.\ D {\bf 55} (1997) 6775,
  [arXiv:hep-ph/9607359].

 \bibitem{witten}
 E.~Witten,  arXiv:hep-ph/0201018;
 M.~Dine, Y.~Nir and Y.~Shadmi,
  Phys.\ Rev.\  D {\bf 66}, 115001 (2002),
  [arXiv:hep-ph/0206268].



\bibitem{so10}
 H. Georgi, in Particle and Fields, AIP, New York (1975), p 575 (C.E. Carlson, Ed);\\
  H.~Fritzsch and P.~Minkowski,
 Ann. Phys. {\bf 93}  (1975) 193.


\bibitem{doublylopsided}
  K.~S.~Babu and S.~M.~Barr, Phys.\ Lett.\ B {\bf 525}
  (2002) 289 [arXiv:hep-ph/0111215];
  Phys.\ Lett.\  B {\bf 381}, 202 (1996),
  [arXiv:hep-ph/9511446].


\bibitem{br}
S.~M.~Barr and S.~Raby,
  Phys.\ Rev.\ Lett.\  {\bf 79} (1997) 4748
  [arXiv:hep-ph/9705366].



\bibitem{slansky}
  R.~Slansky,
  Phys.\ Rept.\  {\bf 79} (1981) 1.



\bibitem{lopsided}
  K.~S.~Babu and S.~M.~Barr,
  Phys.\ Lett.\  B {\bf 381}, 202 (1996),
  [arXiv:hep-ph/9511446];
  C.~H.~Albright and S.~M.~Barr,
  Phys.\ Rev.\  D {\bf 58}, 013002 (1998),
  [arXiv:hep-ph/9712488];
 C.~H.~Albright, K.~S.~Babu and S.~M.~Barr,
  Phys.\ Rev.\ Lett.\  {\bf 81}, 1167 (1998),
  [arXiv:hep-ph/9802314];
   N.~Irges, S.~Lavignac and P.~Ramond,
  Phys.\ Rev.\  D {\bf 58}, 035003 (1998),
  [arXiv:hep-ph/9802334],
   J.~Sato and T.~Yanagida,
  Phys.\ Lett.\  B {\bf 430}, 127 (1998),
  [arXiv:hep-ph/9710516];
   C.~H.~Albright and S.~M.~Barr,
  Phys.\ Lett.\  B {\bf 452}, 287 (1999),
  [arXiv:hep-ph/9901318].





\end{thebibliography}
\end{document}